\documentclass[prb,aps,twocolumn,showpacs]{revtex4-2} 
\usepackage{amsmath,amssymb,amsthm}
\usepackage{graphicx}
\usepackage{subfigure}
\usepackage{amsmath}
\usepackage{amsfonts,amssymb}
\usepackage{bm}
\usepackage{float}
\usepackage{color}
\usepackage{sidecap}
\allowdisplaybreaks
\usepackage{soul}
\setstcolor{red}
\usepackage[normalem]{ulem}
\usepackage{hyperref}
\hypersetup{colorlinks=true,breaklinks,urlcolor=blue,linkcolor=blue,citecolor=blue}

\begin{document}

\title{Tunable non-Hermitian skin effect and topological phases in ladders with staggered nonreciprocal interleg hopping}

\author{Pu-Xuan Li}
\author{Qi-Bo Zeng}
\email{zengqibo@cnu.edu.cn}
\affiliation{Department of Physics, Capital Normal University, Beijing 100048, China}

\begin{abstract}
We investigate the non-Hermitian skin effect (NHSE) and topological phases in a ladder model with staggered nonreciprocal interleg hopping, consisting of a Su-Schrieffer-Heeger chain coupled to a normal tight-binding chain. By tuning the system parameters, we show that the direction of the NHSE under open boundary conditions can be reversed and can even become energy dependent. The NHSE is further characterized by the spectral winding number under periodic boundary conditions. We also demonstrate that the interleg coupling significantly enlarges the topologically nontrivial parameter regime. Remarkably, there are two topological phases characterized by winding numbers $W=\pm1$, with the zero-energy topological edge modes residing on different sites in the two legs. Furthermore, increasing the nonreciprocity of the interleg hopping modifies the topological phase boundaries and eventually drives the system into a topologically trivial phase. Our results establish staggered nonreciprocal interleg hopping as an effective mechanism for engineering both the non-Hermitian skin effect and topological phases in ladder systems.
\end{abstract}
\maketitle
\date{today}

\section{Introduction}
Non-Hermitian systems have become a central topic in condensed matter physics because they provide an effective description of open systems~\cite{Cao2015RMP,Konotop2016RMP,Ashida2020AP,Bergholtz2021RMP}. In contrast to Hermitian systems, non-Hermitian Hamiltonians exhibit a variety of unconventional phenomena, including $\mathcal{PT}$-symmetry breaking~\cite{Bender1998PRL,Bender2007RPP}, exceptional degeneracies~\cite{Heiss2012JPA}, and complex-energy braiding~\cite{Hu2021PRL,Wang2021Nature}, thereby extending topological band theory into genuinely nonequilibrium settings~\cite{Ashida2020AP,Bergholtz2021RMP}. Among the most remarkable discoveries in this field is the non-Hermitian skin effect (NHSE), in which a macroscopic number of eigenstates accumulate exponentially at the system boundaries under open boundary conditions (OBCs)~\cite{Yao2018PRL1}. This phenomenon leads to a profound breakdown of the conventional bulk-boundary correspondence and motivates the development of non-Bloch band theory and the generalized Brillouin zone formalism~\cite{Yao2018PRL1,Yao2018PRL2,YangPRL2020}. Consequently, the NHSE has attracted extensive theoretical and experimental interest~\cite{Alvarez2018PRB,Alvarez2018EPJ,Lee2019PRB,Zhou2019PRB,Kawabata2019PRX,Song2019PRL,Okuma2020PRB,Xiao2020NatPhys,Yoshida2020PRR,Longhi2019PRR,Yi2020PRL,Claes2021PRB,Haga2021PRL,Zeng2022PRB,Zeng2022PRB2,Zeng2023PRA,Xiao2024PRB}. In addition, the energy spectrum of systems exhibiting the NHSE is highly sensitive to boundary conditions~\cite{Xiong2018JPC}, making such systems promising candidates for high-performance quantum sensing~\cite{Budich2020PRL,Koch2022PRR}. The emergence of the NHSE under OBCs is closely related to the existence of point gaps in the complex-energy spectrum under periodic boundary conditions (PBCs)~\cite{Okuma2020PRL,Zhang2020PRL}.

On the other hand, ladder models constitute a natural and versatile extension of one-dimensional lattice systems and have served as an important platform for studying competing hopping processes and topological phenomena in low-dimensional quantum matter~\cite{Liu2012PRB,Li2013NatCom,Sun2016PRA,Ogino2021PRB,Ogino2022PRB,Mondal2023PRB,Parida2024PRB,Downing2024NJP,Aghtouman2024SciRep,Elia2025PRB}. They are particularly relevant for describing layered topological materials and systems coupled to substrates through the proximity effect~\cite{Lu2007PRB,Shoman2015NatCom,Hsieh2016PRL,Cheng2019PRB,Zheng2018PRB}. Compared with a single one-dimensional chain, ladder geometries offer additional degrees of freedom such as interleg hopping for engineering band structures and topological phases, which could substantially modify the properties of the constituent chains and give rise to novel phenomena, including the emergence of flat bands in Creutz ladders~\cite{Sun2017PRB,Kuno2020NJP,Orito2021PRB,Mukherjee2022PRB,Pelegri2024PRB}. Numerous theoretical studies have investigated ladder systems constructed from prototypical one-dimensional topological models, such as the Su-Schrieffer-Heeger (SSH) chain~\cite{Zhang2017PRA,Li2017PRB,Padavic2018PRB,Nersesyan2020PRB,Jangjan2020SciRep,Jangjan2022PRB,Padhan2024PRB,Zeng2026NJP} and the Kitaev chain~\cite{DeGottardi2011NJP,Wu2012PLA,Wakatsuki2014PRB,Maiellaro2018EPJS,Shibata2019PRB,Nehra2020PRR,Xu2024PRB}. More recently, non-Hermitian ladder systems have also attracted considerable attention and have been shown to exhibit a variety of unconventional phenomena~\cite{Yang2018PRB,Liu2019CPB,Islam2022PRB,Wu2022PRA,Mu2022PRB,Liang2022CPB,Qi2023PRB,Wang2023NJP,Chen2024PRA,Jiang2024PRB,Tang2024APL,Tang2025NatCom,Zhang2025PRB,Roy2025PRB,Zhou2025PRB,Wang2025PRB,Ma2025PRB,Roy2025PRR,Sarkar2025arxiv,Lin2025APL,Li2026NatPhys}. For example, Ref.~\cite{Zhou2025PRB} uncovered non-Abelian topology and Ref.~\cite{Li2026NatPhys} revealed the phenomenon of exceptional deficiency by studying different non-Hermitian ladder models composed of two SSH chains. Despite these advances, most previous studies mainly focused on ladder models with nonreciprocal hopping inside each leg, the role of nonreciprocal interleg coupling remains largely unexplored. In particular, it is still unclear how asymmetric interleg hopping influences the NHSE, whether the direction of skin accumulation can be manipulated through the ladder geometry, and how topological phase boundaries evolve in the presence of nonreciprocal interleg coupling.

In this work, we introduce a non-Hermitian ladder model composed of an SSH chain and a uniform tight-binding chain, coupled through staggered nonreciprocal interleg hopping that alternates between odd and even sites. This model is different from the ladder model studied in previous works, in which the nonreciprocal hopping is mainly introduced in each leg. Here both chains are reciprocal by themselves, only the interleg hopping is nonreciprocal. We demonstrate that the staggered nonreciprocity in the interleg hopping induces the NHSE and enables its direction under OBCs to be reversed by tuning the system parameters. Moreover, the NHSE can become energy dependent, such that eigenstates in different energy windows are localized at opposite ends of the ladder. These distinct localization behaviors are characterized by the spectral winding number calculated under PBCs. We also investigate the topological properties of the model and show that the interleg coupling significantly enlarges the topologically nontrivial parameter regime compared with that of an isolated SSH chain. Most interestingly, we find two different topological phases characterized by winding numbers $W=\pm1$, where the corresponding topological zero-energy modes reside on different sites in the two ladder legs. Increasing the nonreciprocity of the interleg hopping modifies the topological phase boundaries and eventually drives the system into a topologically trivial phase once the nonreciprocity exceeds a critical strength. Our work reveals rich NHSE and topological phenomena arising from staggered nonreciprocal interleg coupling and provides a versatile platform for controlling both localization and topology in non-Hermitian ladder systems.

The remainder of this paper is organized as follows. In Sec.~\ref{Sec2}, we introduce the ladder model and give the model Hamiltonian. Section~\ref{Sec3} is devoted to the NHSE in the ladder model. In Sec.~\ref{Sec4}, we investigate the topological phases and the associated edge modes. Finally, we summarize our main results in Sec.~\ref{Sec5}.

\begin{figure}[t]
	\includegraphics[width=3.3in]{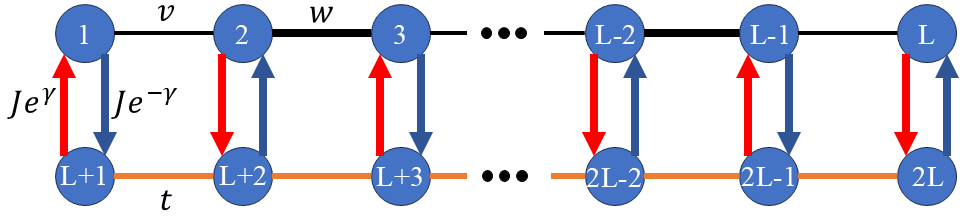}
	\caption{(Color online) Schematic illustration of the ladder model. The upper leg is an SSH chain with alternating hopping amplitudes $v$ and $w$, while the lower leg is a uniform tight-binding chain with hopping amplitude $t$. The interleg hopping is nonreciprocal, with amplitudes $J e^{\gamma}$ and $J e^{-\gamma}$, and alternates between odd and even sites in the two chains.}
	\label{fig1}
\end{figure}

\section{Model Hamiltonian}\label{Sec2}
We consider a ladder model composed of two coupled tight-binding chains, where the upper leg is an SSH chain and the lower leg is a uniform tight-binding chain. The interleg hopping is nonreciprocal and staggered, as illustrated by the red and blue vertical arrows in Fig.~\ref{fig1}. The Hamiltonian of the system is given by
\begin{equation}\label{H}
	H = H_{1} + H_{2} + H_{\perp},
\end{equation}
with
\begin{equation}
	\begin{aligned}
		H_{1} &= \sum_{m=1}^{N} \left(v \hat{c}_{2m-1,1}^\dagger \hat{c}_{2m,1} + w \hat{c}_{2m,1}^\dagger \hat{c}_{2m+1,1} + H.c. \right), \\ 
		H_{2} &= \sum_{m=1}^{N} \left( t \hat{c}_{2m-1,2}^\dagger \hat{c}_{2m,2} + t \hat{c}_{2m,2}^\dagger \hat{c}_{2m+1,2} + H.c. \right), \\
		H_{\perp} &= \sum_{m=1}^{N} \left(J e^{\gamma} \hat{c}_{2m-1,1}^\dagger \hat{c}_{2m-1,2} + J e^{-\gamma} \hat{c}_{2m-1,2}^\dagger \hat{c}_{2m-1,1} \right. \\
		& \qquad \left. + J e^{-\gamma} \hat{c}_{2m,1}^\dagger \hat{c}_{2m,2} + J e^{\gamma} \hat{c}_{2m,2}^\dagger \hat{c}_{2m,1} \right).
	\end{aligned}
\end{equation}
Here, $H_{1}$ describes the SSH chain with alternating intracell and intercell hopping amplitudes $v$ and $w$, respectively, while $H_{2}$ represents the uniform tight-binding chain with nearest-neighbor hopping amplitude $t$. The term $H_{\perp}$ describes the coupling between the two legs. The operators $\hat{c}_{m,\alpha}$ and $\hat{c}_{m,\alpha}^{\dagger}$ ($\alpha=1,2$) are the annihilation and creation operators of spinless fermions at site $m$ on the upper ($\alpha=1$) and lower ($\alpha=2$) legs, respectively. The amplitudes $J e^{\pm\gamma}$ denote the nonreciprocal interleg hopping, whose direction alternates between odd and even sites. Owing to the interleg coupling, each unit cell contains four lattice sites. Throughout this work, the lattice constant is set to unity, $N$ denotes the number of unit cells, and the parameters $v$, $w$, $t$, $J$, and $\gamma$ are all taken to be real. Without loss of generality, we choose $w=1$ as the energy unit.

Figure~\ref{fig1} illustrates the lattice structure of the ladder model together with the site labeling convention adopted throughout this work. For the upper leg, the site indices satisfy $1\le j\le L$, where $L=2N$, whereas the lower leg is labeled by $(L+1)\le j\le 2L$. Accordingly, the real-space Hamiltonian is represented by a $2L\times2L$ matrix. Under OBCs, the energy spectrum is obtained by direct diagonalization of the Hamiltonian matrix.

To characterize the localization properties of the eigenstates and determine the direction of the non-Hermitian skin effect, we employ the directional inverse participation ratio (dIPR), defined as~\cite{Zeng2022PRB}
\begin{equation}
	\text{dIPR}(\psi_n) = \mathcal{P}(\psi_n) \sum_{j=1}^{2L} \frac{|\psi_{n,j}|^4}{\left( \langle \psi_n | \psi_n \rangle \right)^2}, 
\end{equation}
where 
\begin{equation}
	\begin{aligned}
	   \mathcal{P}(\psi_n) =& \mathrm{sgn} \left[ \sum_{j=1}^{L} (j-L/2-\delta) |\psi_{n,j}| \right. \\
	         & \left. + \sum_{j=L+1}^{2L} (j-3L/2-\delta) |\psi_{n,j}|  \right],
	\end{aligned}
\end{equation}
with $\mathrm{sgn}(x)$ denoting the sign function and $\delta$ being a positive constant satisfying $0<\delta<0.5$. $\mathcal{P}(\psi_n)$ extracts the information about whether the eigenstate $\psi_n$ is more localized in the left or right half of the lattice. With the choice of an appropriate value of $\delta$, the reference position $\frac{L}{2}+\delta$ can split each leg into two halves with equal number of lattice sites. For the case we study here, we set $L=100$ in each leg and $\delta=0.2$ such that each of the two halves of the upper and lower legs contains $50$ sites. Here, $|\psi_n\rangle$ is the $n$th eigenstate satisfying the Schr\"odinger equation $H | \psi_n \rangle = E_n |\psi_n \rangle$, where $E_n$ is the corresponding eigenenergy and $\psi_{n,j}$ denotes the $j$th component of $|\psi_n\rangle$. Throughout this work, the eigenstates are ordered according to the real parts of their eigenenergies. For extended states, $\mathrm{dIPR}\rightarrow0$ in the thermodynamic limit, whereas for states localized at the left or right boundary, the dIPR approaches a finite value of order $O(1)$ with a negative or positive sign, respectively. Therefore, the dIPR provides a convenient quantity for distinguishing eigenstates localized at opposite boundaries of the ladder.

\begin{figure}[t]
	\includegraphics[width=3.3in]{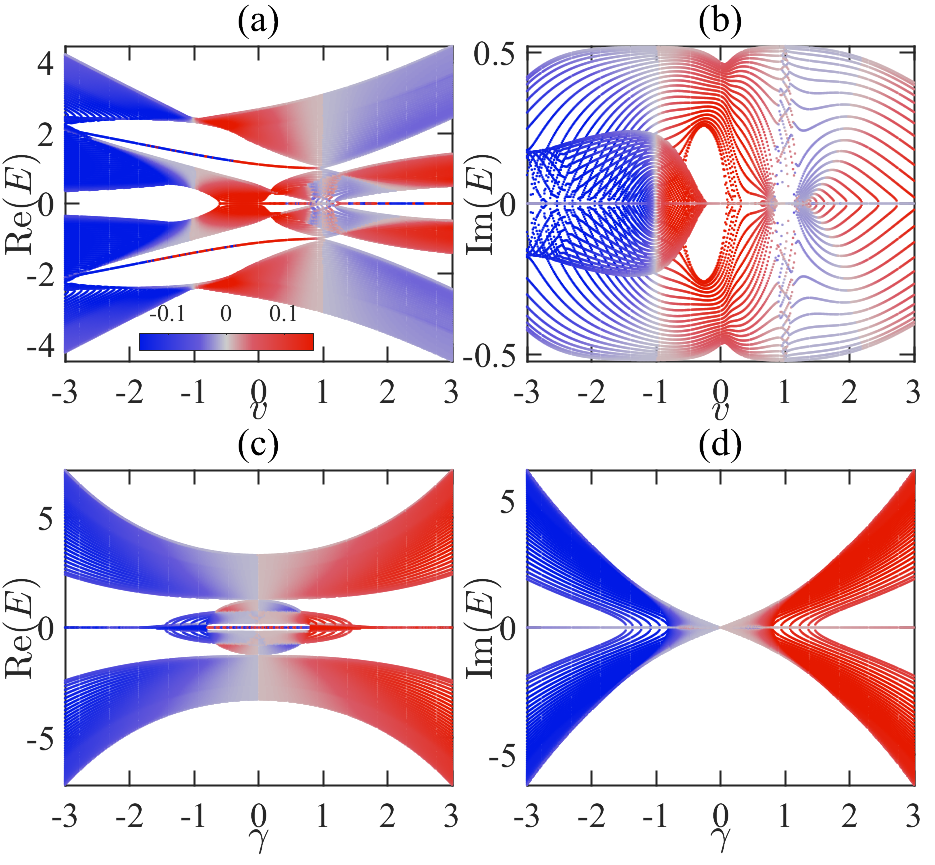}
	\caption{(Color online) Energy spectra of the ladder model under open boundary conditions. (a) and (b) show the real and imaginary parts of the eigenenergies, respectively, as functions of $v$ for $\gamma=-0.5$. The color scale represents the $\mathrm{dIPR}$ of the corresponding eigenstates. Positive (negative) $\mathrm{dIPR}$ values indicate states localized at the right (left) boundary of the ladder, whereas $\mathrm{dIPR}\approx 0$ corresponds to extended states. (c) and (d) show the real and imaginary parts of the eigenenergies, respectively, as functions of $\gamma$ for $v=1.5$. Other parameters are $w=t=J=1$ and $L=100$.}
	\label{fig2}
\end{figure}

\section{Non-Hermitian skin effect}\label{Sec3}
We first investigate the NHSE in the proposed ladder model. If the nonreciprocal interleg hopping is uniform on all sites, no NHSE arises along the ladder direction. In contrast, the staggered nonreciprocal interleg hopping considered here gives rise to the NHSE. More importantly, if one tunes the system parameters, the direction of the NHSE can be reversed and can even become energy dependent. 

\begin{figure}[t]
	\includegraphics[width=3.4in]{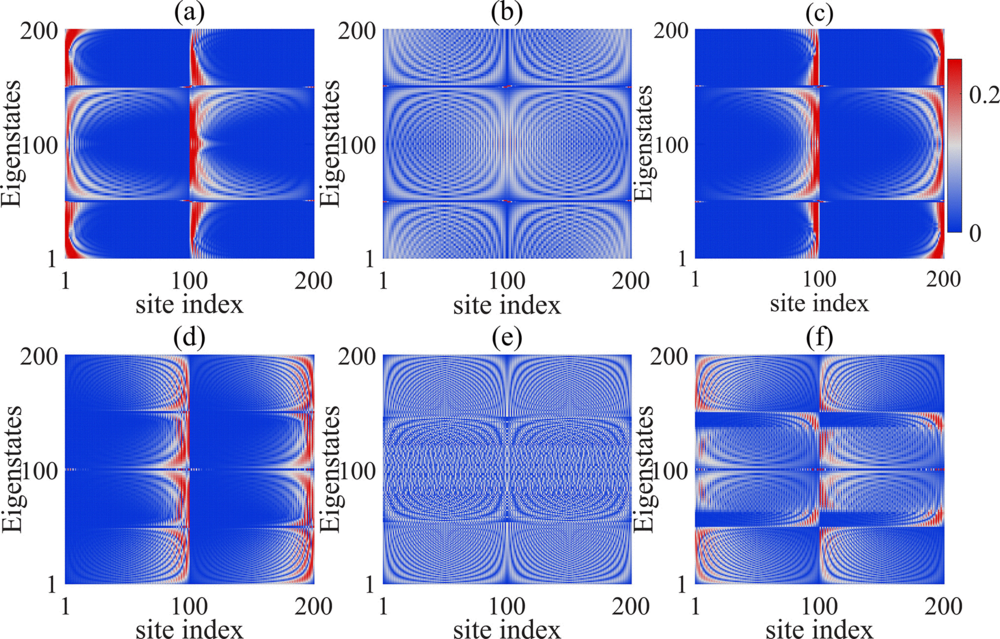}
	\caption{(Color online) Spatial distributions of the eigenstates in the ladder model for different values of $v$: (a) $v=-1.5$, (b) $v=-1.0$, (c) $v=-0.5$, (d) $v=0.5$, (e) $v=1.0$, and (f) $v=1.5$, with $\gamma=-0.5$. The eigenstates are ordered according to the real parts of the corresponding eigenenergies. The color scale represents the amplitude $|\psi|$ of the eigenstates. The site indices $1\leq j\leq100$ correspond to the upper leg of the ladder, whereas $101\leq j\leq200$ correspond to the lower leg. Other parameters are $w=t=J=1$.}
	\label{fig3}
\end{figure}

Figures~\ref{fig2}(a) and \ref{fig2}(b) show the real and imaginary parts of the energy spectrum as functions of $v$ under OBCs, with $w=t=J=1$ and $\gamma=-0.5$. The color scale represents the dIPR of the corresponding eigenstates. For $v<-1$, all dIPR values are negative, indicating that the eigenstates are localized at the left boundary of the ladder. The corresponding spatial distributions are shown in Fig.~\ref{fig3}. As an example, Fig.~\ref{fig3}(a) presents the eigenstate distributions for $v=-1.5$, where all eigenstates are localized near the left boundary. Here, the site indices $1 \leq j \leq 100$ correspond to the upper leg, whereas $101 \leq j \leq 200$ correspond to the lower leg. Therefore, for $v<-1$, the eigenstates are localized at the left boundaries of both legs. At the critical point $v=-1$, the dIPR approaches zero, indicating that the eigenstates become extended and the NHSE disappears. This behavior is confirmed by Fig.~\ref{fig3}(b), where the eigenstates are distributed throughout the entire ladder. As $v$ increases further into the regime $-1<v<1$, the dIPR becomes positive [see Fig. 2(a)], indicating that the direction of the NHSE is reversed. Consequently, all eigenstates become localized at the right boundaries of both legs, as illustrated in Figs. 3(c) and 3(d). Note that near $v=1$, we have states with positive and negative dIPR values, implying that the NHSE becomes energy dependent such that some eigenstates are localized at the left boundary while others are localized at the right boundary.

When $v=1$, the dIPR of all eigenstates again approaches zero, as shown in Figs.~\ref{fig2}(a) and \ref{fig2}(b), indicating that the NHSE disappears and the eigenstates become extended throughout the entire ladder [Fig.~\ref{fig3}(e)]. As $v$ is increased further, the NHSE reemerges with energy dependent: some eigenstates are localized at the left boundary of the ladder, whereas others are localized at the right boundary. This feature becomes more pronounced with increasing $v$, as evidenced by the growing contrast between the positive and negative dIPR values in Fig.~\ref{fig2}(a). The spatial distributions of the eigenstates for $v=1.5$, shown in Fig.~\ref{fig3}(f), further illustrate the energy-dependent nature of the NHSE. In particular, the eigenstates belonging to the lowest and highest energy bands are localized at the left boundary of the ladder, whereas those in the two middle bands are localized at both the left and right boundaries, depending on their energies. Therefore, tuning the parameter $v$ not only reverses the direction of the NHSE but also drives a transition from an energy-independent to an energy-dependent NHSE.

Figures~\ref{fig2}(c) and \ref{fig2}(d) show the real and imaginary parts of the energy spectrum as functions of $\gamma$, which controls the nonreciprocity of the interleg hopping. As the nonreciprocity is varied, the NHSE is correspondingly tuned, and an energy-dependent NHSE emerges in the regime $|\gamma|<1$. As $|\gamma|$ increases, the energy-dependent behavior gradually disappears. For sufficiently large positive (negative) values of $\gamma$, all eigenstates become localized at the right (left) boundary of the ladder.

It is well known that the emergence of the NHSE under OBCs is intimately related to the point-gap topology of the PBC energy spectrum. Under PBCs, the Hamiltonian in Eq.~(\ref{H}) can be transformed into momentum space via a Fourier transformation, yielding the Bloch Hamiltonian
\begin{equation}\label{Hk}
	H(k)= \begin{pmatrix}
		0 & v+we^{-ik} & Je^{\gamma} & 0 \\
		v+we^{ik} & 0 & 0 & Je^{-\gamma} \\
		Je^{-\gamma} & 0 & 0 & t+te^{-ik} \\
		0 & Je^{\gamma} & t+te^{ik} & 0
	\end{pmatrix}.
\end{equation}
Diagonalizing $H(k)$ yields four energy bands,
\begin{equation}\label{Ek}
	\begin{aligned}
		E(k) =& \pm \sqrt{A(k) \pm B(k)}, 
	\end{aligned}
\end{equation}
where
\begin{equation}
	\begin{aligned}
		A(k) =& \frac{1}{2} \left[ 2J^{2} + |p|^2 + |q|^2 \right], \\
		B(k) =& \sqrt{ A(k)^{2} - C(k) },
	\end{aligned}
\end{equation}
with $p=v+we^{ik}, q=t+te^{ik}$, and
\begin{equation}
	C(k)= |p|^2 |q|^2+ J^{4} - J^{2} e^{-2\gamma} p^* q - J^{2} e^{2\gamma} p q^*.
\end{equation}

\begin{figure}[t]
	\includegraphics[width=3.4in]{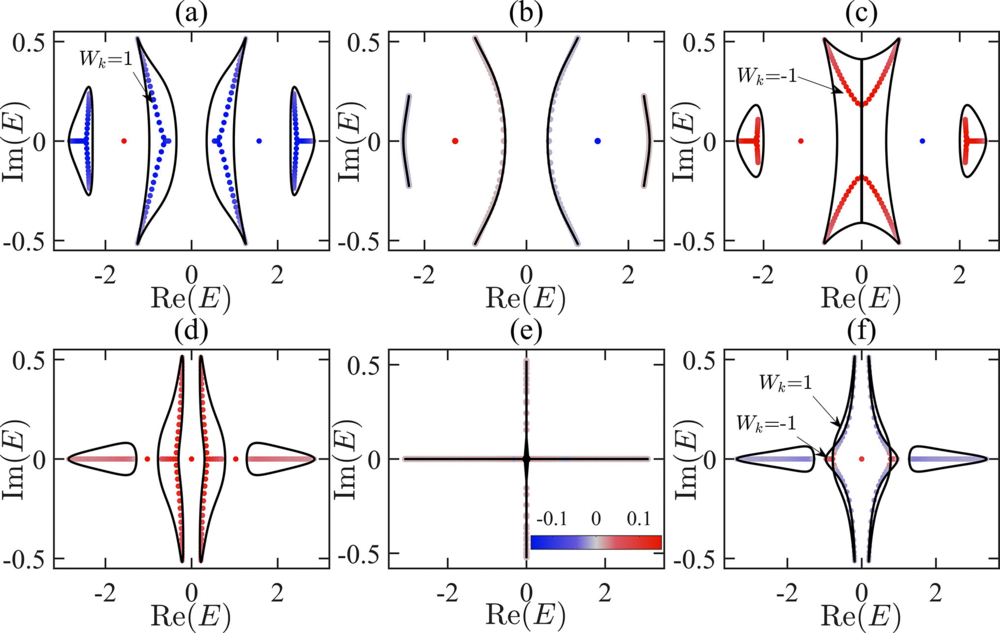}
	\caption{(Color online) Energy spectra of the ladder model under OBCs (colored dots) and PBCs (black dots) for $\gamma=-0.5$ and different values of $v$: (a) $v=-1.5$, (b) $v=-1.0$, (c) $v=-0.5$, (d) $v=0.5$, (e) $v=1.0$, and (f) $v=1.5$. The color scale represents the $\mathrm{dIPR}$ of the OBC eigenstates. The labels $W_k=\pm1$ denote the spectral winding numbers of the corresponding PBC spectra, which determine the direction of the NHSE under OBCs. Other parameters are the same as those in Fig.~\ref{fig3}.}
	\label{fig4}
\end{figure}

Figure~\ref{fig4} compares the energy spectra under OBCs and PBCs for the ladder model with $v$ varying from $-1.5$ to $1.5$, while the other parameters are the same as those in Fig.~\ref{fig3}. As shown in Fig.~\ref{fig4}, the OBC spectrum is enclosed by the corresponding PBC spectrum except at the critical points $v=\pm1$ [Figs.~\ref{fig4}(b) and \ref{fig4}(e)]. At these two critical points, the OBC and PBC spectra coincide and form open arcs rather than closed loops in the complex-energy plane. Consequently, the NHSE is absent under OBCs, consistent with the extended eigenstates shown in Figs.~\ref{fig3}(b) and \ref{fig3}(e). For the remaining parameter regimes, the direction of the NHSE can be characterized by the spectral winding number of the PBC spectrum, defined as
\begin{equation}\label{Wk}
	\begin{aligned}
		W_k =& \frac{1}{2\pi i} \int_{-\pi}^{\pi} dk\, \partial_k \ln \det \left[ H(k)-E_B \right].
	\end{aligned}
\end{equation}
Here, $E_B$ is a reference energy, which is typically chosen inside the PBC spectral loop. A spectral winding number of $W_k=+1$ corresponds to eigenstates localized at the left boundary of the ladder, as illustrated in Fig.~\ref{fig4}(a), whereas $W_k=-1$ indicates localization at the right boundary [Figs.~\ref{fig4}(c) and \ref{fig4}(d)]. In both cases, the NHSE is energy independent, and all eigenstates are localized at the same boundary under OBCs. In contrast, for $v=1.5$ [Fig.~\ref{fig4}(f)], the two central bands develop self-intersections, and the spectral winding number changes sign across different energy regions. Consequently, eigenstates with different energies become localized at opposite boundaries of the ladder, indicating the emergence of an energy-dependent NHSE. 

The critical points at $v=\pm1$ can be understood by analyzing the PBC spectra, since the corresponding energy bands do not form closed loops in the complex-energy plane. We first consider the case $v=1$, corresponding to $v=t=w$ in Eq.~(\ref{Ek}). In this case, $C(k)$ reduces to
\begin{equation}
	\begin{aligned}
		C(k) =&|q|^4 + J^{4}- J^{2} e^{-2\gamma} |q|^2 - J^{2} e^{2\gamma} |q|^2 \\
		=& J^{4} + 4w^4 ( 1 +  \cos{k} )^2 \\
		&- 2J^{2} w^2 ( 1 + 1 \cos{k} ) (e^{-2\gamma} + e^{2\gamma}),
	\end{aligned}
\end{equation}
which yields
\begin{equation}
	\begin{aligned}
	E^2 (k) &= A(k) \pm B(k) \\
	&= J^2 + 4w^2 \cos^2 \frac{k}{2}  \pm 4Jw \left| \cos \frac{k}{2} \right| \cosh \gamma.
	\end{aligned}
\end{equation}
The above expression is either nonnegative or nonpositive for all $k$, depending on the value of $\gamma$. Consequently, $E(k)$ is either purely real or purely imaginary and therefore remains on the real or imaginary axis of the complex-energy plane. As a result, the PBC spectrum cannot form closed loops, consistent with the numerical results shown in Fig.~\ref{fig4}(e). Accordingly, the NHSE is absent at $v=1$, and all eigenstates remain extended under OBCs.

For the other critical point, $v=-1$, setting $-v=t=w$ yields
\begin{equation}
	\begin{aligned}
		E^2 (k) &= J^2 +2w^2 \\
		&\pm 2 \sqrt{J^2w^2 + w^4\cos^2k + i J^2 w^2 \sin k \sinh (2\gamma)},
	\end{aligned}
\end{equation}
which likewise does not form closed loops in the complex-energy plane. Consequently, the PBC spectrum does not exhibit a point gap, consistent with the numerical results shown in Fig.~\ref{fig4}(b). Therefore, the NHSE is also absent at $v=-1$.

For the energy-dependent NHSE shown in Fig.~\ref{fig2}(f), we can see the critical energies separate the bulk eigenstates localized at opposite boundaries of the ladder. These critical energies correspond to the self-intersection points of the PBC spectra and are referred to as NHSE edges~\cite{Zeng2022PRB2}. For the ladder model considered here, the NHSE edges can be determined numerically from the PBC spectra, since analytical expressions are generally unavailable due to the complexity of the spectral structure.

From another perspective, the ladder model we study here can be transformed into a one-dimensional chain by relabeling the lattice sites, where nonreciprocal and long-range hopping exist in the system. Due to the interplay between the nonreciprocal hopping and the long-range hopping, the PBC spectral topology will become more complex than the model with only nearest-neighboring nonreciprocal hopping (e.g., see Ref.~\cite{Zeng2022PRB2}). From the intertwined structures of the PBC spectra (see Fig.~\ref{fig4}), we can observe the phenomena of reversal NHSE and the energy-dependent NHSE. Compared with the ladder models studied in previous works, which mainly focus on the nonreciprocity inside each leg, our model demonstrates that the staggered longitude nonreciprocal hopping between the two legs can influence the behaviors of the NHSE along the transverse direction. So the staggered nonreciprocal interleg hopping provides a different way to tune the NHSE.

\section{Topological phase and edge modes}\label{Sec4}
\begin{figure}[t]
	\includegraphics[width=3.4in]{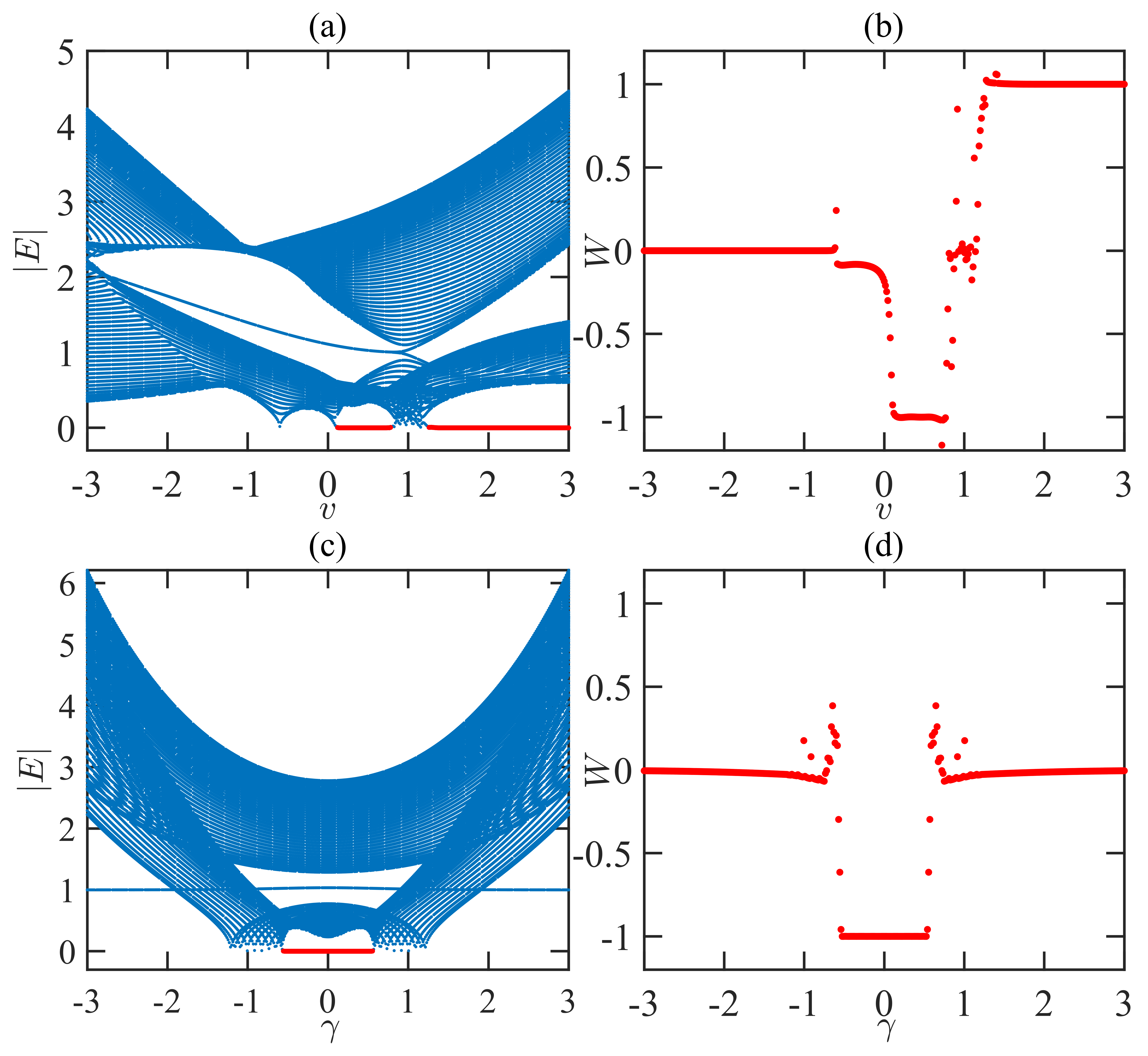}
	\caption{(Color online) (a) Absolute values of the eigenenergies as functions of $v$ for $\gamma=-0.5$. The zero-energy modes are highlighted by red dots. (b) Real-space winding number as a function of $v$. (c) Absolute values of the eigenenergies as functions of $\gamma$ for $v=0.5$; (d) the corresponding real-space winding number. Other parameters are $J=t=w=1$ and $L=100$.}
	\label{fig5}
\end{figure}

\begin{figure}[t]
	\includegraphics[width=3.4in]{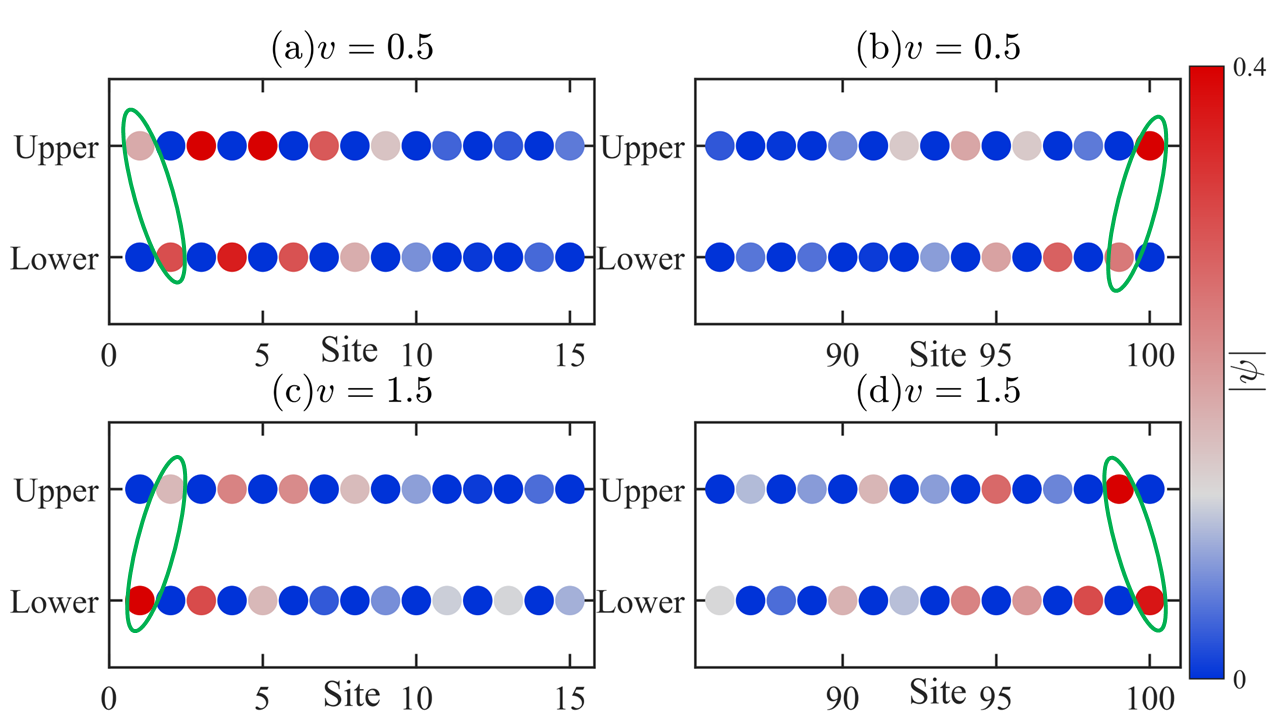}
	\caption{(Color online) Spatial distributions of the topological zero-energy edge modes in the ladder model. (a) and (b) show the left and right edge modes, respectively, for the system with $v=0.5$. (c) and (d) show the corresponding edge modes in system with $v=1.5$. The dots represent the lattice sites of the ladder model, and the color of each site indicates the magnitude of the wave function component, as shown by the colorbar. The green ellipse marks the even and odd sites occupied by the edge modes near the boundary of each leg. Other parameters are $J=t=w=1$ and $L=100$.}
	\label{fig6}
\end{figure}

We next investigate the topological phases of the ladder model. Since the upper leg is an SSH chain, which hosts topological zero-energy edge modes in its topologically nontrivial phase, one naturally expects the ladder model to support topological edge states. Indeed, the OBC spectra shown in Figs.~\ref{fig3} and \ref{fig4} already reveal the emergence of both zero-energy and finite-energy in-gap states. To highlight the zero-energy modes, Fig.~\ref{fig5}(a) shows the absolute values of the eigenenergies as a function of $v$ for $J=t=w=1$ and $\gamma=-0.5$, where the zero-energy modes are marked by red dots. Two distinct parameter regions supporting zero-energy modes are identified: $0.1 \lesssim v \lesssim 0.8$ and $v \gtrsim 1.26$. These are the topologically nontrivial regimes. As illustrated in Fig.~\ref{fig6}, these zero-energy modes are localized at the boundaries of the ladder and therefore correspond to topological edge modes. 

Comparing our model with the conventional single SSH chain, where the nontrivial phase with zero-energy edge modes exists only in the region with $|v|<|w|$, we can see that the parameter region for the existence of zero-energy edge modes is largely extended. Thus the nontrivial regime is extensively enhanced in our ladder model due to the coupling of the SSH chain with another chain. By tuning the interleg hopping terms, we can effectively tune the topological phases in the ladder.

To characterize the topological phases, we calculate the real-space winding number under open boundary conditions~\cite{Song2019PRL}. The Hamiltonian preserves chiral symmetry,
\begin{equation}
	SHS^{-1}=-H,
\end{equation} 
with $S=\mathcal{I}\otimes\sigma_z$, where $\mathcal{I}$ is the identity matrix and $\sigma_z$ is the Pauli matrix. Let $|\psi_{nR}^{\pm}\rangle$ and $|\psi_{nL}^{\pm}\rangle$ denote the biorthonormal right and left eigenstates, which satisfy
\begin{equation}
	\begin{aligned}
		H | \psi^{\pm}_{nR} \rangle &= \pm E_n | \psi^{\pm}_{nR} \rangle, \\
		H^\dagger | \psi^{\pm}_{nL} \rangle &= \pm E_n^\star | \psi^{\pm}_{nL} \rangle,
	\end{aligned}
\end{equation}
respectively. Excluding the edge modes, the open-boundary $Q$ matrix is defined as
\begin{equation}
	Q = \sum_n \left( | \psi^{+}_{nR} \rangle \langle \psi^{+}_{nL} | - | \psi^{-}_{nR} \rangle \langle \psi^{-}_{nL} | \right),
\end{equation}
and the corresponding real-space winding number is given by
\begin{equation}
	W = \frac{1}{2L^\prime} \mathrm{Tr^\prime} \left( SQ [Q, X] \right),
\end{equation} 
where $X$ is the position operator and $\mathrm{Tr}^\prime$ denotes the trace over a central region of length $L^\prime$. In our calculations, the trace region $L^\prime$ is evaluated over the central $50\%$ of the system in order to suppress boundary contributions. For $L=100$, the trace is evaluated over the $51$st to the $150$th eigenstates, which are sorted according to the real parts of the eigenenergies. To exclude the zero-energy boundary states from the $Q$-matrix construction, we retain only eigenstates satisfying $|E|>0.005$. The winding number takes the quantized values of $W=\pm1$ in the topologically nontrivial phases and $W=0$ in the trivial phase. Figure~\ref{fig5}(b) shows the real-space winding number as a function of $v$. The regions with $W=\pm1$ coincide precisely with those supporting zero-energy edge modes in Fig.~\ref{fig5}(a). With increasing system size, the quantization of the winding number will become clearer and the transition around the critical points will be sharper (see the Appendix).

We further examine the energy spectrum and the real-space winding number as functions of $\gamma$ in Figs.~\ref{fig5}(c) and \ref{fig5}(d) for the system with $v=0.5$. As the nonreciprocity of the interleg hopping increases, the topological phase gradually shrinks. Once the nonreciprocity exceeds a critical strength, the zero-energy edge modes disappear, accompanied by the real-space winding number changing from $W=\pm1$ to $W=0$, indicating a topological phase transition to the trivial phase.

In Fig. 6, we show the spatial distributions of the topological zero-energy modes in the ladder model. The dots represent the lattice sites of the ladder model, while the color of each site indicates the magnitude of the wave function component. Figures~\ref{fig6}(a) and \ref{fig6}(b) demonstrate the distributions of the two zero-energy modes in the $W=-1$ phase. For the zero-energy mode localized at the left end of the ladder, the wave function is distributed over the odd sites in the upper leg and the even sites in the lower leg near the left end, as indicated by the sites encircled by the green ellipse in Fig.~\ref{fig6}(a). However, for the zero-energy mode localized at the right end, the wave function resides on the even sites of the upper leg and the odd sites of the lower leg, which are  encircled by the green ellipse in Fig.~\ref{fig6}(b). On the contrary, for the topological regime with $W=+1$, the distribution of the left zero-energy mode is localized on the even (odd) sites of the upper (lower) leg, while the right edge mode distributes on the odd (even) sites in the upper (lower) leg, as shown in Figs.~\ref{fig6}(c) and \ref{fig6}(d). The distinction between the two topological phases is therefore reflected in the different distribution patterns of the topological edge modes, which gives a physical explanation distinguishing the $W=-1$ and $W=+1$ topological phases.

\begin{figure}[t]
	\includegraphics[width=3.2in]{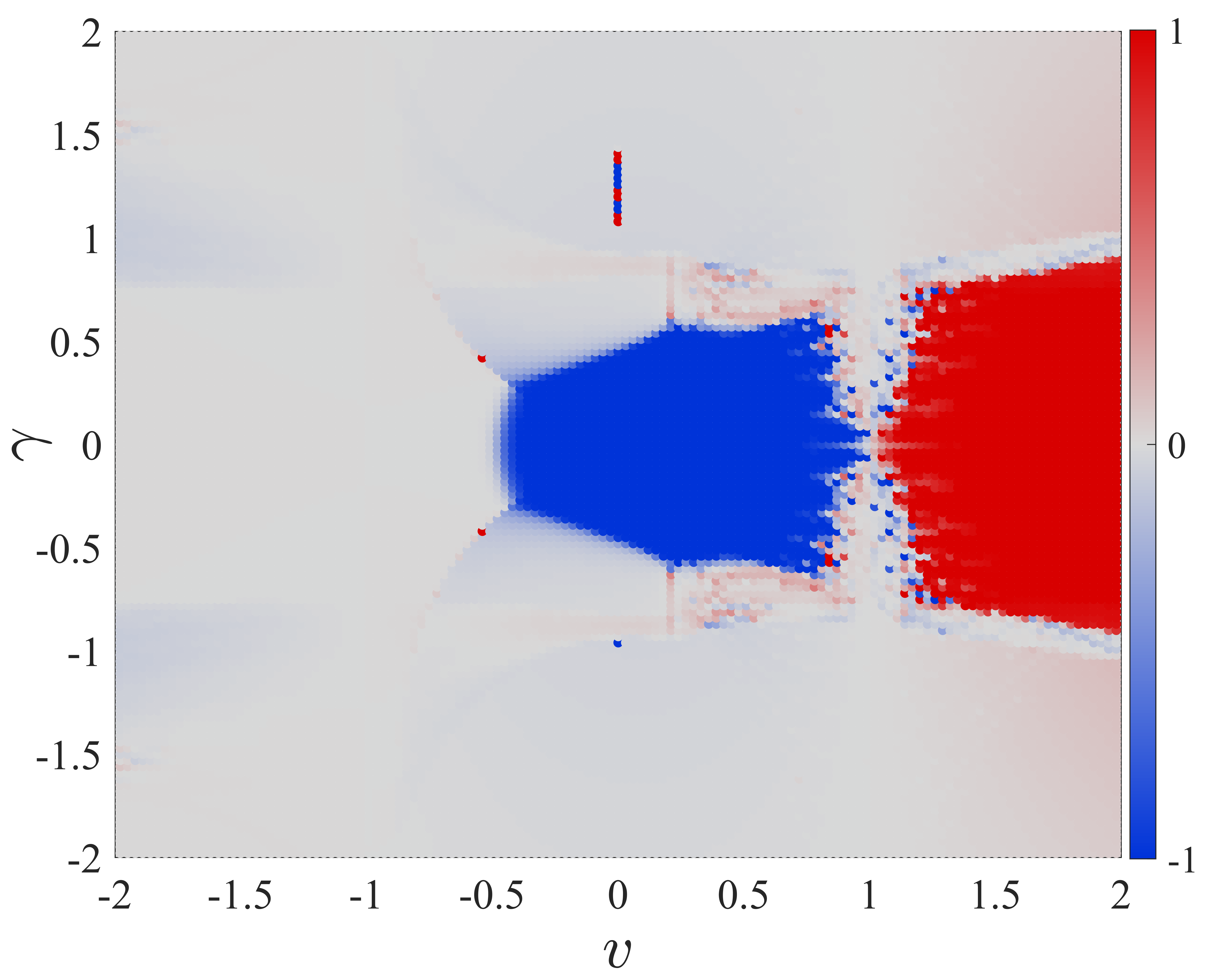}
	\caption{(Color online) Topological phase diagram of the ladder model in the $v$--$\gamma$ plane for $J=t=w=1$ and $L=100$. The blue region corresponds to the topological phase with $W=-1$, while the red region represents the topological phase with $W=+1$.} 
	\label{fig7}
\end{figure}

We further present the phase diagram of the ladder model in Fig.~\ref{fig7}, obtained by calculating the real-space winding number for different values of $v$ and $\gamma$. The blue region corresponds to the topological phase with $W=-1$, while the red region represents the phase with $W=+1$. It is evident that the interleg coupling substantially enlarges the topologically nontrivial parameter regime compared with that of an isolated SSH chain. Moreover, from the phase diagram we can also see that increasing the nonreciprocity in the interleg hopping will finally destroy the nontrivial phase. Note that near the phase boundary separating the $W=-1$ and $W=+1$ phases, a few isolated blue and red points appear. These points originate from the fact that the bulk bands are not well separated in this parameter region, as shown in Figs.~\ref{fig5}(a) and \ref{fig5}(c). Consequently, the real-space winding number is not well defined, making the numerical calculation more susceptible to finite-precision errors.

In addition to the zero-energy edge modes, finite-energy edge modes appear inside the band gaps, as shown in Fig.~\ref{fig5}(a). Moreover, their energies remain nearly independent of $\gamma$, as illustrated in Fig.~\ref{fig5}(c). Unlike the zero-energy edge modes, these finite-energy edge modes are not topologically protected. Instead, they originate from the coupling between the SSH chain and the normal chain, and can be understood within the framework of perturbation theory~\cite{Zeng2026NJP}. Moreover, the energies of these finite-energy in-gap modes vary continuously with the model parameters, implying that they are not topologically protected like the topological zero-energy edge modes.

From Figs.~\ref{fig2} and \ref{fig5}, we can see that the parameter regimes for topologically nontrivial phases and the regimes for different NHSE behaviros are not the same. For the NHSE, we find that $v=-1$ ($v=+1$) is the critical point where the NHSE direction reverse (become energy dependent) for the system with $\gamma=-0.5$. As for the topological phase, we have no phase transition around $v=-1$, the topological phase transition happens around $v=+1$. These results further unveil the different origins of the NHSE and the topological phase transitions.

\section{Summary}\label{Sec5}
In summary, we investigated the non-Hermitian skin effect and topological phases of a non-Hermitian ladder model consisting of an SSH chain coupled to a normal tight-binding chain through staggered nonreciprocal interleg hopping. We showed that the NHSE can be effectively tuned by varying the system parameters, leading to a reversal of the skin accumulation direction or even an energy-dependent NHSE. The different NHSE regimes are characterized by the spectral winding number of the PBC spectrum, and the corresponding critical points are identified analytically. We also demonstrated that the interleg coupling enlarges the topologically nontrivial regime compared with that of an isolated SSH chain. Furthermore, the topological zero-energy edge modes were found to reside on different sites in the two legs of the ladder and are characterized by the real-space winding numbers $W=-1$ and $W=+1$. Our results reveal the rich interplay between nonreciprocity, the NHSE, and topology in non-Hermitian ladder systems, and provide a simple platform for engineering tunable skin effects and topological phases in quasi-one-dimensional lattices.

As to the experimental realizations, the one-dimensional SSH model has been experimentally realized in various platforms, including ultracold atoms \cite{Atala2013NatPhys,Lohse2016NatPhys,Nakajima2016NatPhys}, photonic systems \cite{Wang2009Nature}, and acoustic metamaterials \cite{Susstrunk2015Science}. Our ladder model can, in principle, be realized by extending such platforms to include an additional tight-binding chain and staggered interleg couplings. In particular, the reciprocal intracell and intercell hoppings can be implemented through the corresponding couplings between neighboring sites, while the nonreciprocal interleg hoppings can be engineered using suitable active elements or controlled coupling schemes. In the past few years, topolectrical circuits composed of resistors, inductors and capacitors have been widely used to simulate various topological and non-Hermitian systems \cite{Lee2018CommunPhys,Hofmann2019PRL,Zhang2025PRB,Jin2025PRL}. Such circuits are relatively accessible to implement and can be readily scaled, providing a flexible platform for realizing lattice models with complex and nonreciprocal couplings. Within this framework, the circuit Laplacian can be designed to realize an effective circuit representation of our ladder Hamiltonian, with the staggered nonreciprocal interleg hoppings implemented using appropriate active circuit elements, such as negative-impedance converters with current inversion. The spectral properties and boundary localization can then be investigated through frequency-dependent impedance measurements, together with the corresponding spectral and eigenmode analysis of the circuit Laplacian. Therefore, topolectrical circuits provide a feasible platform for experimentally probing the NHSE, the topological phases, and the zero-energy modes of the non-Hermitian ladder model studied in this work.

\begin{acknowledgments}
	This work is supported by the National Natural Science Foundation of China under Grant No. 12204326.
\end{acknowledgments}

\begin{figure}[t]
	\includegraphics[width=3.4in]{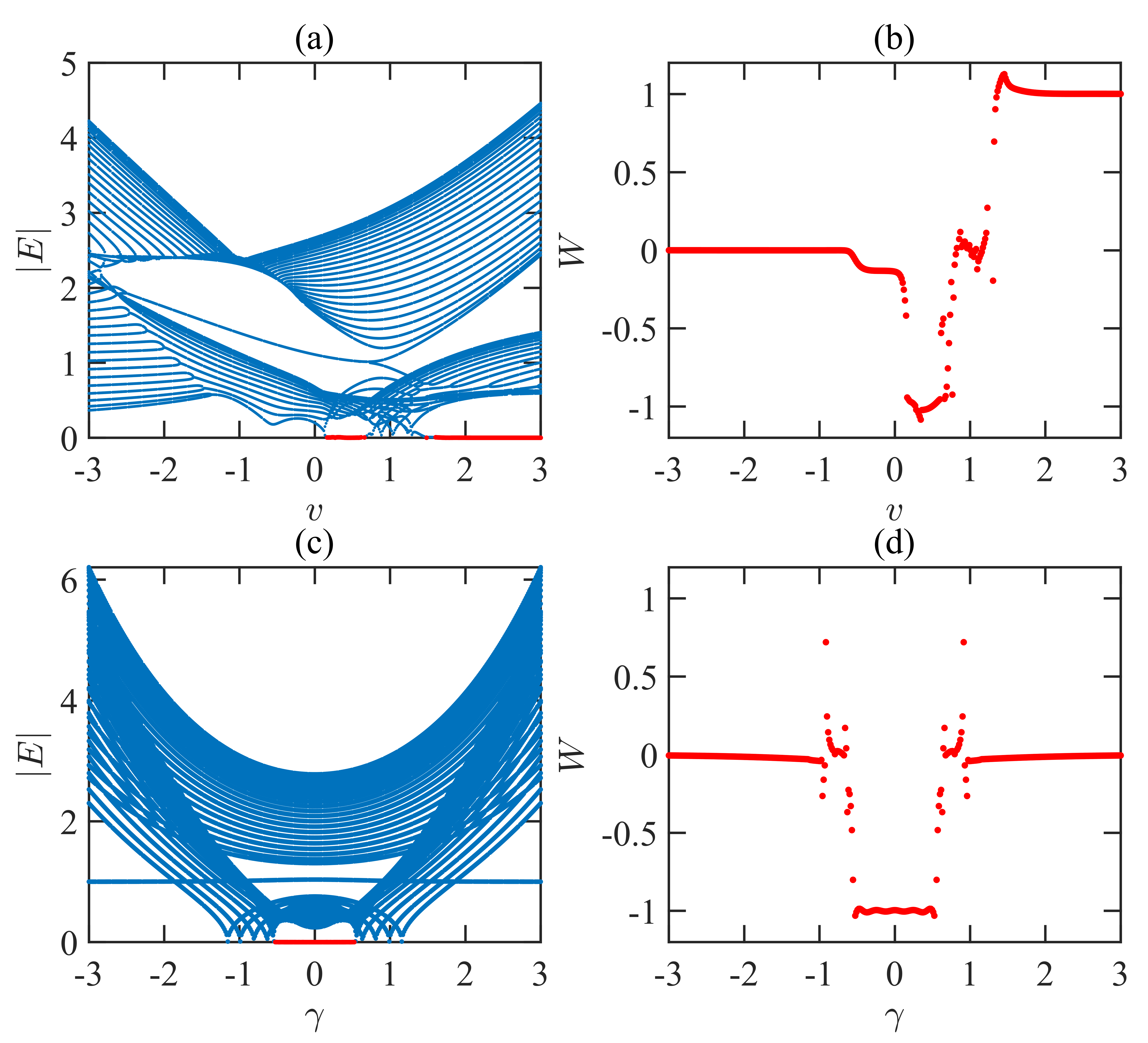}
	\caption{(a) Absolute values of the eigenenergies as a function $v$ for $\gamma=-0.5$; the zero-energy modes are represented by red dots. (b) Real-space winding number as a function of $v$. (c) and (d) The spectrum and the winding number as a function of $\gamma$ when $v=0.5$, respectively. Here the system size is $L=50$. Other parameters are $J=t=w=1$.} 
	\label{figA1}
\end{figure}

\begin{figure}[t]
	\includegraphics[width=3.4in]{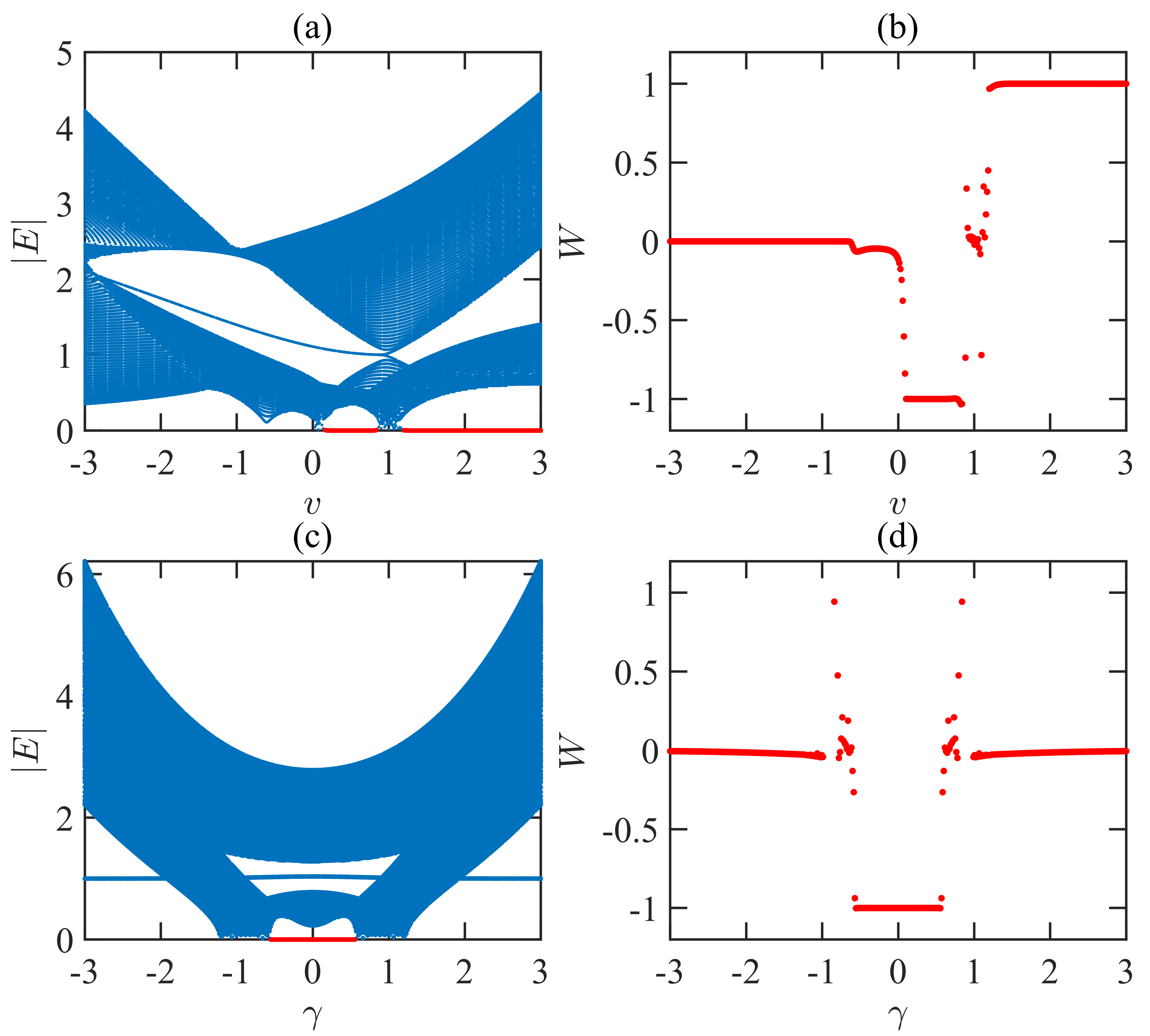}
	\caption{Eigenenergy spectra and the real-space winding number as a function of $v$ or $\gamma$. All the parameters are the same as in Fig.~\ref{figA1}, but the system size is $L=150$.}
	\label{figA2}
\end{figure}

\setcounter{equation}{0}
\renewcommand*{\theequation}{A\arabic{equation}}

\section*{Appendix}\label{Appendix}
In this appendix, we provide more numerical results for the calculation of the real-space winding number, which is used to characterize the topological phase in the ladder model introduced in this paper. To examine the finite-size stability of the zero-energy boundary states and the quantization of the open-boundary bulk topological invariant, we calculated the energy spectra and the winding numbers for the system with $L=50$ and $L=150$ see Figs.~\ref{figA1} and \ref{figA2}. The system parameters are the same as in Fig.~\ref{fig5}. We can see that as the system size increases, the finite-size deviations of the calculated real-space winding number become smaller, and the numerical values approach the expected quantized values more closely. The plateaus at $W=0$, $\pm 1$ become more stable. On the other hand, the regime with zero-energy modes in the spectrum become clearer and the transition gets sharper near the phase transition critical points.

\end{document}